\documentclass{article}

\usepackage{arxiv}

\usepackage[utf8]{inputenc} 
\usepackage[T1]{fontenc}    
\usepackage{hyperref}       
\usepackage{url}            
\usepackage{booktabs}       
\usepackage{amsfonts}       
\usepackage{nicefrac}       
\usepackage{microtype}      
\usepackage{lipsum}		
\usepackage{graphicx}
\usepackage{natbib}
\usepackage{doi}

\usepackage{amsmath}

\title{Measurement and applications of position bias in a marketplace search engine}

\author{ {\hspace{1mm}Richard Demsyn-Jones} \\
	Thumbtack \\
	\texttt{demsynjones@thumbtack.com} \\
	\texttt{rdemsynjones@gmail.com}
}

\renewcommand{\headeright}{}
\renewcommand{\undertitle}{}

\hypersetup{
pdftitle={Measurement and applications of position bias in a marketplace search engine},
pdfauthor={Richard Demsyn-Jones},
pdfkeywords={position bias, search ranking, learning to rank, inverse propensity scoring, propensity estimation},
}

\begin{document}
\maketitle

\begin{abstract}
	Search engines intentionally influence user behavior by picking and ranking the list of results. Users engage with the highest results both because of their prominent placement and because they are typically the most relevant documents. Search engine ranking algorithms need to identify relevance while incorporating the influence of the search engine itself. This paper describes our efforts at Thumbtack to understand the impact of ranking, including the empirical results of a randomization program. In the context of a consumer marketplace we discuss practical details of model choice, experiment design, bias calculation, and machine learning model adaptation. We include a novel discussion of how ranking bias may not only affect labels, but also model features. The randomization program led to improved models, motivated internal scenario analysis, and enabled user-facing scenario tooling.
\end{abstract}


\section{Introduction}

Free text search bars are central parts of some of our most successful consumer technologies, from informational search engines like Google to marketplaces like Amazon. Search systems seek to present the most relevant results most visibly to their users. By doing so they take advantage of, and entrench, top-to-bottom examination behavior. The selection and ranking of search results is an important area of research. We verified the importance of our ranking through this work, and we also believe that search engine ranking is non-controversially important and widely regarded as a key determinant in the success of Google and other search-based services \citep{Joachims07}.

There is a large literature examining and modeling consumer behavior with search engines \citep{Chapelle09, Dupret08, Yue10, Agarwal19}. Often the goal is to create ideal rankings that are not not biased by the endogenous history of prior ranking \citep{Joachims17, Wang18, Aslanyan19, Agarwal19b, Vardasbi20, Ovaisi20}. In this paper we describe a path of applied research in one marketplace, where we continually experiment with new ranking algorithms and invest in data for subsequent iterations, such that we can improve our ranking by making it easier for customers to match with professionals who can do their jobs. We initially built algorithms with biased features and loss functions until investing in a randomization program to identify bias from ranking. We applied our research from the randomization program into two improved models, first successfully supplanting our baseline model with a new model using unbiased features, before subsequently supplanting that model with a model using an unbiased estimator. Both of those iterations not only showed improvements in offline model metrics, but also improved our aggregate marketplace metrics as part of a wider ranking system. Our bias estimates also proved useful beyond improved ranking, through internal hypothesis evaluation and external tooling.

The main contributions of this work are:

\begin{itemize}
    \item Position bias adjustments for features, with or without adjustments for position bias in model optimization
    \item Discussion of power and cost trade-offs in randomized allocation schemes for measuring position bias
    \item Experimental validation of RandPair position bias factors for an Inverse Propensity Scoring (IPS) model outside of Google's internal datasets and experiments for Gmail and Drive
    \item Sharing a simple calculation for position bias factors from search logs with randomized allocation
\end{itemize}

\section{Thumbtack search}
\label{sec:thumbtack_search}

Thumbtack is a large marketplace for professional services. The Thumbtack marketplace has \textit{visitors} (or \textit{customers} or \textit{searchers}) who have projects for which they need professional help, and \textit{professionals} who seek customers for their services. A typical Thumbtack visitor might be a homeowner who uses Thumbtack to seek professional help with fixing, maintaining, or improving their home. Thumbtack hosts a wide variety of services, including house cleaning, plumbing, wedding planning, dog walking, tax planning, and many others. The professionals on Thumbtack may be self-employed individuals, small local businesses, or large national providers. They may specialize in specific services or have many employees across a range of services.

\begin{figure}[ht]
  \centering
  \includegraphics[width=350px]{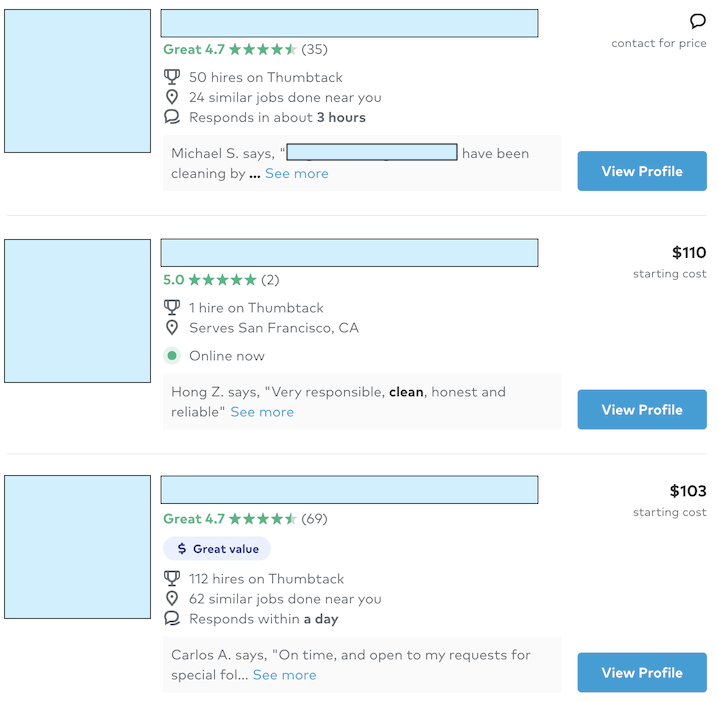}
  \caption{Example of search results on Thumbtack, with identifying information hidden.}
  \label{fig:example_professional_list}
\end{figure}

Finding the right professional for a project can be daunting. Professionals are not interchangeable. They have different skills, locations, prices, and other attributes that may matter to customers. Thumbtack aims to facilitate successful matches in the marketplace, where a customer and a professional both want to undertake a project with each other. We provide a search engine, allowing customers to specify their needs through text search or click navigation. Visitors see a search results lists of applicable professionals, such as the one shown in Figure \ref{fig:example_professional_list}. They can narrow down their search with filters on project requirements and scheduling, and they can view detailed profiles of individual professionals.

Thumbtack’s search engine is central to the customer experience. While some customers can find the right professional through directly reaching their profile or by providing project details and asking Thumbtack to find professionals on their behalf, the majority of visitors see vertically listed search results, peruse summary information or the profiles of the professionals shown, and then select from them. Sometimes we use \textit{pagination}, where we show 10 professionals at a time, while other interfaces may instead contain a continually scrolling list.

The result at the top of the first page is the 1st position (indexing from 1), with the subsequent result the 2nd position, and so forth. A “higher” position means higher visually to a user, with a lower numeric index, such that (for example) position 1 is the highest position since it is the very first listing when observed top-to-bottom.

\begin{figure}[ht]
  \centering
  \includegraphics[width=350px]{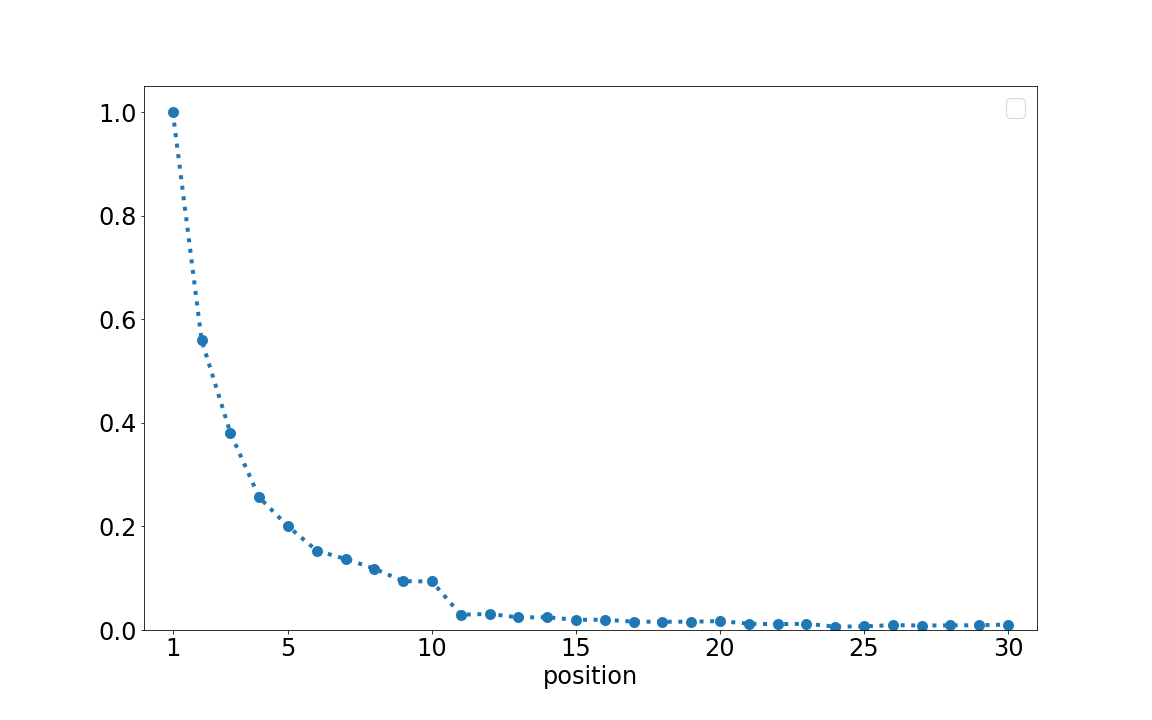}
  \caption{Relative contact rate by position, for searches with 30 or more professionals available.}
  \label{fig:contacts_by_position}
\end{figure}

As shown in Figure \ref{fig:contacts_by_position}, we have diminishing contact rates by position, except for slight reversals at the bottom of pages. Diminishing contact rates could be caused by a combination of customer preference for higher results and our success at ranking better matches at the top of our search results.

Thumbtack’s ideal ranking is that which facilitates the highest likelihood of the best match. Following the Probability Ranking Principle that "documents should be ranked in order of the probability of relevance or usefulness" \citep{Singh18}, we maximize ranking utility by ranking professionals in descending order of our predicted relevance to the customer \citep{Robertson77}. We want pairings of customers and professionals that will choose to work together, both better off than if they had not collaborated, and collectively better off than if any other pairing had matched. When a customer and professional connect and the professional solves the customer’s need we call that a \textit{job done}.

\section{Related work}

\subsection{Terminology and background}

The term \textit{position bias} is sometimes used to refer to any causal relationship between rank and outcomes \citep{Yue10, Ovaisi20}. In other papers it is used to refer to a specific model of such causal relationships \citep{Li20, Wang18, Vardasbi20}. We will use the term \textit{ranking bias} to refer to any form of causal impact of ranking on outcomes, to distinguish clearly between these two uses of position bias in the literature.

Much of the literature describes \textit{clicks} as a metric of interest \citep{Yue10, Joachims17}. Some search engines observe clicks with high accuracy and have limited visibility on any further action. A click on a Thumbtack search result does not lead away from Thumbtack to outside of our visibility. Some downstream activity stays on our platform, creating other metrics of match success. In this paper we will focus on \textit{contact}, which is when a customer selects a professional, answers a series of standard questions about their project, and submits those answers along with a message to the professional. The professional will then see the project listed alongside other contacts and can respond to the customer to coordinate on the project.

\textit{Examination} describes a typically unobserved process whereby customers look at and consider a search result \citep{Joachims17, Wang18, Agarwal19}. With pagination of search results we can know that all search results on non-visited pages were not examined. At Thumbtack we track scrolling in the user interface, so we can further identify professionals that were not examined. When a user scrolls such that a search result is fully shown in their interface we call that result \textit{viewed}. View is only an approximation for examination, since a visitor may scroll rapidly to see several professionals but only truly consider some of them.

The literature uses the term \textit{query} to refer to the context in which a ranked search result is generated. In many search engines, the query is a text input. The concept generalizes to other available information about the searcher. A \textit{document} is an item ranked for the visitor. The visitor has a goal of finding a relevant document that satisfies the purpose of their search.

In the context of Thumbtack, the query may be text from the search bar or it may be another mechanism of reaching a search result, such as navigation through recommended searches. Documents correspond to professionals, examination corresponds to a serious consideration of the professional, and relevance corresponds to a professional that will lead to a mutually beneficial match between the customer and the professional. 

\subsubsection{The cascade model}

In a cascade model visitors view the search results from top to bottom, considering each result in turn, and they stop when they find a relevant result \citep{Craswell08, Dupret08}.

The cascade model in its most literal form is easy to refute. The model assumes users stop exactly when they find a relevant result or exhaust the search results. We have searches where visitors did not contact any professionals, yet did not view the entire list of professionals we returned. A notable minority of our searches have multiple contacts, indicating that the customer did not halt their search at the first relevant professional. Customers do not always contact the last viewed search result. On the contrary, even in searches with substantial scrolling we still see contacts strongly diminish at lower positions. In the cascade model the final examined search result is always the selected one, so even with noise in view events we would expect an inverse relationship between position and contact when customers have scrolled. We also observe backwards scroll behavior, where in a minority of searches we see backtracking of at least a few positions.

\subsubsection{User Browsing Model (UBM) and Dynamic Bayesian Network (DBN)}

In the UBM, users scroll top-to-bottom, examine only some of the results, can select multiple results, and are more likely to cease their search as relevant results become further apart \citep{Dupret08}.

In the DBN, users scroll top-to-bottom, examining each item, can click multiple items, and will stop when they click an item that is satisfying \citep{Chapelle09}.

Those two models were published by Yahoo researchers during the same time period, and they share a very similar framing. They articulate more complex behavior than the simpler cascade model.

In their literal formation they cannot perfectly fit Thumbtack data, since they have exclusively forward movement while Thumbtack has some reverse movement. Not only do customers change the direction of their scanning, but we also see contacts happening in reverse order. In the case where customers make exactly two contacts from the same search result list, about 30\% of the time they contacted the lower ranked professional first.

\subsubsection{The position-based propensity model (PBM)}

In the position-based propensity model (or position bias model), the probability of contact equals the probability of relevance multiplied by the probability of examination, where the probability of examination is only a function of position \citep{Joachims17, Wang18}. Using the variables defined in Table \ref{table:symbols}, the PBM is summarized in Equations \ref{eq:pbm_assumption} and \ref{eq:pbm_base}.

\begin{table}
  \centering
  \caption{Symbols used in this text}
  \begin{tabular}{ccl}
    \toprule
    Symbol&Meaning\\
    \midrule
    $q$ & Query (or search)\\
    $d$ & Document (or professional)\\
    $C$ & Click (or contact)\\
    $y$ & Set of documents and positions for a query\\
    $k$ & Position\\
    $E$ & Examination, Expectation\\
    $R$ & Relevance\\
    $\theta_k$ & P(E = 1 | position = k)\\
    $r$ & Estimate of $\theta$ from logs\\
    $n$ & Sample count\\
    $f$ & Ranking algorithm\\
    $\Delta$ & Loss function\\
    $\lambda$ & A weight function within the loss function\\
  \bottomrule
\end{tabular}
\label{table:symbols}
\end{table}

\begin{equation}
  C = 1 \Leftrightarrow E = 1, R = 1
  \label{eq:pbm_assumption}
\end{equation}

\begin{equation}
  P(C = 1|q,d,k) = P(R = 1|q,d) \cdot P(E = 1|k)
  \label{eq:pbm_base}
\end{equation}

The isolation of position impact into a probability of examination is the \textit{separability hypothesis} \citep{Dupret08}, which implies none of the quality-of-context bias from \citet{Joachims07}. Each position has an independent likelihood of examination, as if the visitor generates a random number for each position to decide whether to examine the corresponding document. The PBM has been extensively used because of its simplicity and has a successful track record in applications \citep{Agarwal19b}. The PBM is a useful model, despite lacking the more realistic behavioral aspects of the trust bias model, the UBM, or the DBN.

In the quantification of position bias, a higher position bias means that a position is more likely to receive contacts, holding all else equal, such that position 1 typically has the largest position bias. We can describe relative position bias between any two positions, such that one position is given a position bias of 1 and the other position has a position bias that indicates their interaction rate relative to the reference position. Typically position 1 is the reference position, and we describe a full set of position biases relative to it. We assign the value of 1 to the first position, and then the number for position 2 is the ratio of expected rate of contacts for a professional if shown in position 2 to the expected contacts for the same professional if shown in position 1. Position 1 could have a value of 1, position 2 a value of 0.6, position 3 a value of 0.4, and so forth. This is the common quantification used in the position bias literature.

\subsection{Extensions}

The click model literature is rich with alternative models. The \textit{trust bias} extension to the PBM removes the separability hypothesis. It supports the nuance that we only observe perceived relevance, which depends not only on actual relevance but also position \citep{Agarwal19b, Vardasbi20}. Other approaches support multiple document modalities \citep{Chen12, WangChao2013}, unify different forms of bias \citep{Yi21}, support non-click successful outcomes in the mobile ranking \citep{Mao18}, add personalized bias estimations for users with substantial history \citep{ZhangJunqi22}, or consider iterative comparisons of neighbouring documents \citep{ZhangRuizhe21}.

\section{Modeling position bias at Thumbtack}

\subsection{Choice of the PBM}

The literature makes the distinction between navigational queries where searchers know what document they want and use the search engine as a tool to find the right link, and informational queries where users seek information more broadly and multiple results could be relevant \citep{Dupret08, Chapelle09}. A third goal that Thumbtack users could have is project fulfillment, where they have a specific task and seek to to hire a single professional or compare multiple professionals. This is uncommon in the literature. Although some papers come from retailers like eBay \citep{Aslanyan19} or TripAdvisor \citep{Li20}, many papers come from partnerships with navigational and informational search engines like those from Google \citep{Agarwal19, Agarwal19b, Qin20, Yue10, Wang18}, Microsoft \citep{Joachims07, Guo09, Ling17}, and Yahoo \citep{Chapelle09, Dupret08}.

The PBM is flexible enough to fit most observed Thumbtack searcher behavior. Users may arrive and browse our search results top to bottom. They could stop when they have found a sufficiently relevant professional, or when the results have become demonstrably less relevant (as in the UBM), or when they have exhausted the attention they were willing to commit to the search process. They could then review the set they have examined and contact the professional(s) with the highest observable relevance, possibly with some preference for higher ranked results due to trust in the ranking system. Multiple contacts can be explained by the limited relevance of individual professionals, the project’s urgency and its need for a backup plan, or informational searches. Actual user behavior could vary on any of those aspects, and our observed data could be a mix of many types of searchers.

We assume that position bias is constant for any quality level and for any search query on Thumbtack, but varies by the type of interaction (profile view, contact, and so forth) and the type of product interface used, motivated by similar concerns as the attribute-based propensity framing of \citet{Qin20}.

\section{Measuring position bias}

\subsection{Alternatives to randomization}

\begin{figure}[ht]
  \centering
  \includegraphics[width=350px]{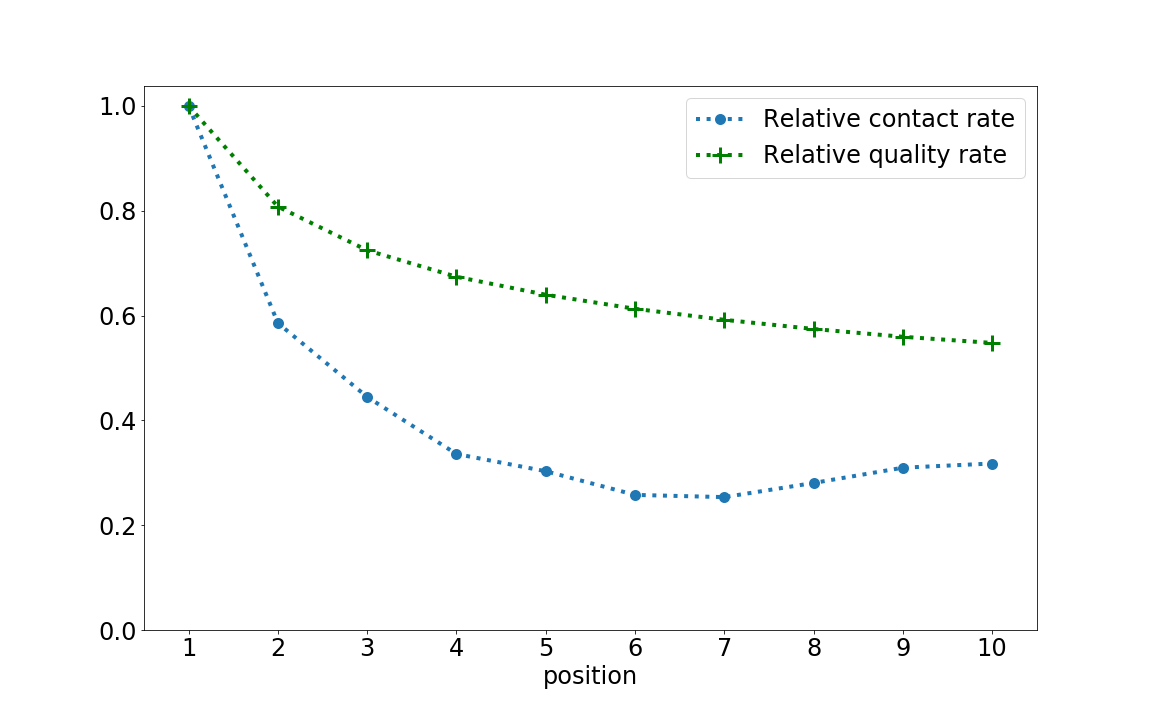}
  \caption{Relative rates when 10 or more professionals are viewed.}
  \label{fig:contacts_vs_quality}
\end{figure}

Since we have scroll data we could assume that observed scrolls represent examination. However, we don’t believe that observation in its most literal sense explains the full extent of ranking bias in our marketplace. Consider the sharper drop in contact rates than quality metrics even when all professionals are viewed, as shown in Figure \ref{fig:contacts_vs_quality} for one of our internal estimates of quality. In the PBM, examination by position is unconditional on examination of other positions, and when two items are both observed their click probabilities vary only on relevance.

We can try to measure ranking bias through regression. Conceptually, we can include all quality factors, trying to “control” for them, and additionally include position as a feature \citep{Ling17, Wang18, Haldar20}. The position coefficients represent position bias. The concept of using position as a feature to separate its impact extends to other model types such as decision trees and neural networks \citep{Ling17}. One issue with this approach is omitted variable bias: if there exist quality factors that are correlated with position but not included in the regression, then their effect could be misattributed to the effect of position. Similarly, the effect of position can be attributed to other features, such that when "there is a high correlation between them, the attribution can be arbitrary" \citep{Wang18}. We could also try an instrumental variables (IV) approach, using a two-stage regression where the first stage predicts position and the second stage uses predicted position as a coefficient. Critically, the first stage prediction needs to accurately predict position without having the same correlation with unobserved quality factors. Another method for unbiased ranking with non-random data is to use Expectations-Maximization (EM) algorithms \citep{Wang18}.

Those approaches are promising but are not without risk of inaccurate position bias estimation. Position as a feature performed poorly when evaluated in \citet{Wang18}. The EM approach had mixed evidence when proposed in \citet{Wang18}, performed poorly in \citet{Aslanyan19}, and was challenged in \citet{Agarwal19} with the explanation that “Defining an accurate relevance model is just as difficult as the learning-to-rank problem itself, and a misspecified relevance model can lead to biased propensity estimates”.

Thumbtack search has considerable regular variation due to ranking experiments, where we continually iterate to help customers match with professionals more easily. This is similar to the context in \citet{Agarwal19}. Their approach treats variation due to ranking experiments as "virtual swap interventions", which also sounds promising for Thumbtack's use case.

We decided that randomization methods had little chance for error, were typically precise while requiring limited sample relative to our scale, and would be highly convincing internally due to their simplicity and lack of modeler choice parameters.

\subsection{Randomization for position bias}

Fully randomizing the entire search results for some subset of searches could lead to a direct estimate of position bias \citep{Wang18}. Over enough observations, the average match quality of professionals would be constant across positions. Measuring the advantage of the first position over the second position would be as simple as measuring the rate of contacts to professionals in those two positions across all searches that had at least two professionals. The downside of any randomized algorithm is degradation of the search results, coming at a cost to successful matches. Full randomization would lead to significant degradation, hurting market participants during the randomization period and possibly leading to damage to Thumbtack's reputation due to visibly subpar search results. 

\begin{figure}[ht]
  \centering
  \includegraphics[width=350px]{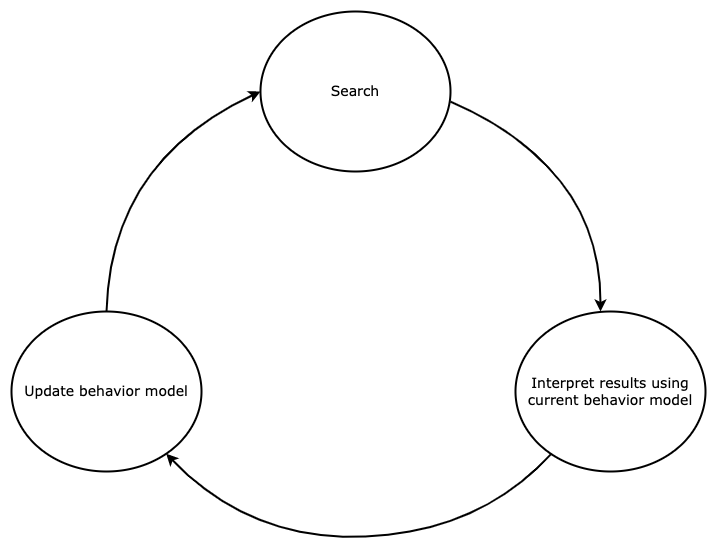}
  \caption{Behavior adaptation model.}
  \label{fig:behavior_states}
\end{figure}

Papers in the ranking bias literature typically assume a fixed behavior model. We propose a behavior adaptation model such that visitor behavior models change in response to ranking, as shown in Figure \ref{fig:behavior_states}. In this adaptation model, visitors choose a behavior model to maximize their utility subject to their expectation about how the ranking algorithm works. As they observe the ranking they adjust their expectation about the ranking algorithm.

This implies some risk that customer behavior is fundamentally different with a random list than with an intelligently ordered list. Perhaps with an intelligently ordered list searchers may trust in our ranking to some extent, leading to our present position bias. When viewing an entirely random list, the unintuitive ordering may be observable to customers, and they may discard this trust. The degree of position bias may change when visitors recognize a poor ranking algorithm. In an extreme case they may change their algorithm from a classic behavior model to one of equally examining a fixed number of professionals.

Instead of fully random search results, we implemented the RandPair algorithm described in \citet{Wang18}. RandPair randomly swaps a single pair of adjacent professionals in each treated search result. While this still comes at some cost, it should be much less significant than the full randomization algorithm. To the extent that there is any quality-of-context bias in search results \citep{Joachims07}, such that the entire ordering matters when customers browse, RandPair involves less disturbance than full randomization. Additionally, estimates from RandPair can be used later to validate other approaches, as in \citet{LiRuilin20}.

\subsection{Power calculations}

We did not know ahead of time how much sample we needed, since "sample size calculation requires assumptions that typically cannot really be tested until the data have been collected" \citep{Gelman06}. Power calculations for experiments are often computed using an effect size (or minimum detectable effect) that is meaningful to decision-making based on the experiment, "the smallest difference or effect that the researcher considers to be clinically relevant" \citep{Pancholi09}. For this randomization program we did not know what level of precision for bias estimates for each position would lead to meaningful changes in the quality of our ranking or for our other use cases.

We decided to emphasize neighbouring positions rather than effects relative to the 1st position so that we could better discern between professionals who we show near each other in search results. These are professionals that could reasonably be reordered by more informed models, in contrast to how the quality gap between the first listed result and most lower results may be large enough that better measurement of position bias is unlikely to lead to a reversal in relative ordering. We wanted to be able to measure the position bias between any neighbouring positions such that if position bias between the two existed then we had a high likelihood of measuring bias factors with statistically significance differences.

We can place a likely upper bound on our effect sizes by using observed contacts rates by position. Observed rates embed both the effects of position bias and of quality, both of which have the same sign of correlation with respect to position. For example, position 1 has the highest position bias and (ideally) the highest average match quality, both of which should lead to higher contact rates in position 1 than other positions. As such, the effect of position bias itself should be smaller than observed interaction rates. We decided to compute power calculations for any neighbouring positions under the hypothesis that half of our observed contact difference is due to position bias. We thought this was a conservative policy for sample estimates, since the decay in contact rates exceeded the decay in estimated match quality, as shown earlier in Figure \ref{fig:contacts_vs_quality}.

We measured position bias separately for several interfaces on our website. For example, after customers peruse the search results, select a professional, and contact them with project details, we then identify and present three more professionals who specifically match those project details. The user interface is substantially different, with the three professionals shown horizontally in a smaller grid overlapping the search results. We estimated power for position bias estimates for this type of list and for others that vary from our typical search results.

Decays in contact rates are steep in the first few positions, with position 1 receiving far more contacts than position 2, position 2 still noticeably more than position 3, and so forth at diminishing rates, as shown earlier in Figure \ref{fig:contacts_by_position}. Position biases among these positions, if accounting for a meaningful proportion of the observed contact rate differences, would be very easy to measure with statistical significance. The position-contact curve is not always smooth, first due to non-monotonicity at the bottom of pages (e.g. position 10 receives more contacts than position 9), secondly due to very steep drops in interactions at page boundaries (e.g. position 10 receiving far more interaction than position 11). Within the second and third pages of search results we observe contact rates close enough and events sparse enough that we see inconsistent average rates over moderate samples. Our power calculations demonstrated that we would not be able to accurately measure position bias in a reasonable time frame between neighbouring positions this deep into the search results.

\subsection{Traffic allocation}

We first isolated the randomization program to 50\% of our traffic. This means that 50\% of traffic definitely had no randomization, while the other 50\% of traffic was subject to the randomization program and typically had a pair of swapped professionals. Randomizing into 50\% of traffic allowed us to very easily and reliably track the aggregate cost of the randomization program using our internal tools for A/B experiments. 

We decided to have adjacent swaps for each pair up through positions 10 and 11 where we expected relative comparisons to be statistically measurable. Additionally, we included a swap of positions 11 and 19 to give us an estimate of the total decay over the second page, which we could attribute to various positions in a manner consistent with the tactic of "interpolate between estimates at well-chosen ranks and/or employ smoothing" from \citet{Joachims17}. The final swap is positions 11 and 19 because we expect position 20 at the bottom of the second page to benefit from position bias relative to position 19. Understanding the net impact between positions 11 and 20 would not help us understand the slope between position 11 and the monotonically worse positions below it.

For the proportion of traffic allocated to the randomization program, with uniform probability we selected a pair to swap from our 11 total swap candidates (positions 1 and 2 through positions 11 and 19) and then performed the swap if the search results were long enough to contain both positions. Table \ref{table:sampleallocation} describes our traffic allocation.

\begin{table}
  \centering
  \caption{Traffic allocation with 50\% holdout and 11 swap permutations}
  \begin{tabular}{ccl}
    \toprule
    Non-stochastic results & 50\%\\
    Intend to swap positions 1 and 2 & 50\% / 11 = 4.55\%\\
    ... & \\
    Intend to swap positions 10 and 11 & 50\% / 11 = 4.55\%\\
    Intend to swap positions 11 and 19 & 50\% / 11 = 4.55\%\\
    \midrule
    \textbf{Total Thumbtack searchers} & \textbf{100\%}\\
  \bottomrule
\end{tabular}
\label{table:sampleallocation}
\end{table}

\subsection{Calculation from search logs}

If we swapped positions $k$ and $k + 1$ in 50\% of our traffic, the expected aggregate quality of professionals in each position would be balanced in the limit. Instead we swap any adjacent pair much less often. Consider an example of positions 8 and 9. Since each position is the chosen swap 4.55\% of the time, then the average match quality at position 8 is still very close to the non-randomized quality of position 8 professionals, and the quality difference between position 8 and position 9 will be roughly similar to when there is was no randomization. 

For comparison of positions $k$ and $k^\prime$, where $k^\prime > k$ (specifically $k^\prime = k + 1$ in our program), consider a corpus of searches that have at least $k^\prime$ professionals. We can only perform a random swap if both the higher and lower position are occupied for the particular search. A list with $k$ professionals might have slightly different behavior than a list with $k + 1$ or more. Professionals on shorter lists may have a higher likelihood of interaction (holding quality equal) due to fewer competitors and on average more prior filtering from visitors.

To construct a quality-neutral estimate for position $k$ in its comparison with position $k^\prime$, which will will call $r_{k,k^\prime}^k$, we take an unweighted average of the interaction rates of the natural position $k$ professionals and the professionals swapped up from position $k^\prime$ to position $k$. This is shown in Equation \ref{eq:rand_1}, with $C$ for contacts and $n$ for sample counts, where we use the first superscript to indicate the original position and the second superscript to indicate final position. For our position $k^\prime$ quality-neutral estimate we take an unweighted average of the interaction rates for professionals naturally ranked in position $k^\prime$ and those moved down from position $k$ to position $k^\prime$, as in Equation \ref{eq:rand_2}. 

\begin{equation}
  r_{k,k^\prime}^k = 0.5 \cdot \left[ \frac{C^{k,k}}{n^{k,k}} + \frac{C^{k^\prime,k}}{n^{k^\prime,k}} \right]
  \label{eq:rand_1}
\end{equation}

\begin{equation}
  r_{k,k^\prime}^{k^\prime} = 0.5 \cdot \left[ \frac{C^{k^\prime,k^\prime}}{n^{k^\prime,k^\prime}} + \frac{C^{k,k^\prime}}{n^{k,k^\prime}} \right]
  \label{eq:rand_2}
\end{equation}

Multiplying by half is for interpretability and will be omitted for convenience going forward. The progression in Equation \ref{eq:rand_3} shows how the expected value of $r_{k,k^\prime}^k$ is our position bias factor times the additive expected relevance of the documents naturally at position $k$ and those naturally at position $k^\prime$.

\begin{equation}
\begin{aligned}
E[r_{k,k^\prime}^k] & = \frac{E[C^{k,k}]}{n^{k,k}} + \frac{E[C^{k^\prime,k}]}{n^{k^\prime,k}} \\
& = \frac{\sum_{i=1}^{n^{k,k}} \theta_k R(d_i^{k,k})}{n^{k,k}} +  \frac{\sum_{i=1}^{n^{k^\prime,k}} \theta_k R(d_i^{k^\prime,k})}{n^{k^\prime,k}} \\
& = \theta_k[E[R(d^{k,k})] + E[R(d^{k^\prime,k})]]
  \label{eq:rand_3}
  \end{aligned}
\end{equation}

Note that relevance is only used to determine the original position and not the random selection, such that $R(d^{k,k^\prime}) = R(d^{k,k})$ and $R(d^{k^\prime,k}) = R(d^{k^\prime,k^\prime})$. Substituting these equivalences and dividing by the same derivation for $E[r_{k,k^\prime}^{k^\prime}]$ leads us to Equation \ref{eq:rand_4}. This shows that the relative ratios of these quality-neutral click rates are unbiased estimators of our relative propensity.

\begin{equation}
\begin{aligned}
\frac{E[r_{k,k^\prime}^k]}{E[r_{k,k^\prime}^{k^\prime}]} = \frac{\theta_k}{\theta_{k^\prime}} \cdot \frac{E[R(d^{k,k})] + E[R(d^{k^\prime,k^\prime})]}{E[R(d^{k^\prime,k^\prime})] + E[R(d^{k,k})]} = \frac{\theta_k}{\theta_{k^\prime}} 
  \label{eq:rand_4}
  \end{aligned}
\end{equation}

This formulation uses data from all contacts to construct position bias estimates of minimal variance. This is conceptually similar to the description in \citet{Agarwal19} of weighing each observation by how often it was shown in each position.

This gives us relative position bias estimates for every pair of adjacent positions. We place them all on the same scale, relative to position 1, by assuming transitivity through the search results. The position 1 versus 3 ratio is calculated by multiplying the relative position bias of 1 versus 2 with the relative position bias of 2 versus 3, as described in \citet{Wang18}.

\subsection{Outcome}

We ran our randomization program over an extended time frame. Aggregate cost was tolerable, with small mean effects on contacts and bidirectional matches as shown in Table \ref{table:programcost}. Our most impactful exchange was the first two positions, and that only occurred in approximately 1/11th of the sample where we had randomization. We found large and monotonic position bias, except for where we expected non-monotonicity at the bottom of the first page of search results. Notably, the position bias was much less substantial than the estimates we used from earlier non-randomized data, implying that we had been overvaluing professionals at low positions for their contacts and undervaluing professionals at high positions for theirs.

While some papers stress the "bookkeeping overhead" of tracking randomization \citep{Joachims17, Aslanyan19}, we needed only minimal added tracking. We also had no difficulty implementing the position bias estimates in SQL from search logs. This allowed us to have graphs of the position bias estimates updated automatically as we accumulated more data, with the results visible to any observers in the company. This transparency helped communicate and document the method.

\begin{table}
  \centering
  \caption{Cost of the randomization program relative to the non-randomized baseline sample}
  \begin{tabular}{ccl}
    \toprule
    Metric&Performance&p-value\\
    \midrule
    Rate of visitors with >= 1 contact & -0.47\% & 0.09\\
    Rate of visitors with >= 1 match & -0.46\% & 0.14\\
  \bottomrule
\end{tabular}
\label{table:programcost}
\end{table}

\begin{figure}[ht]
  \centering
  \includegraphics[width=350px]{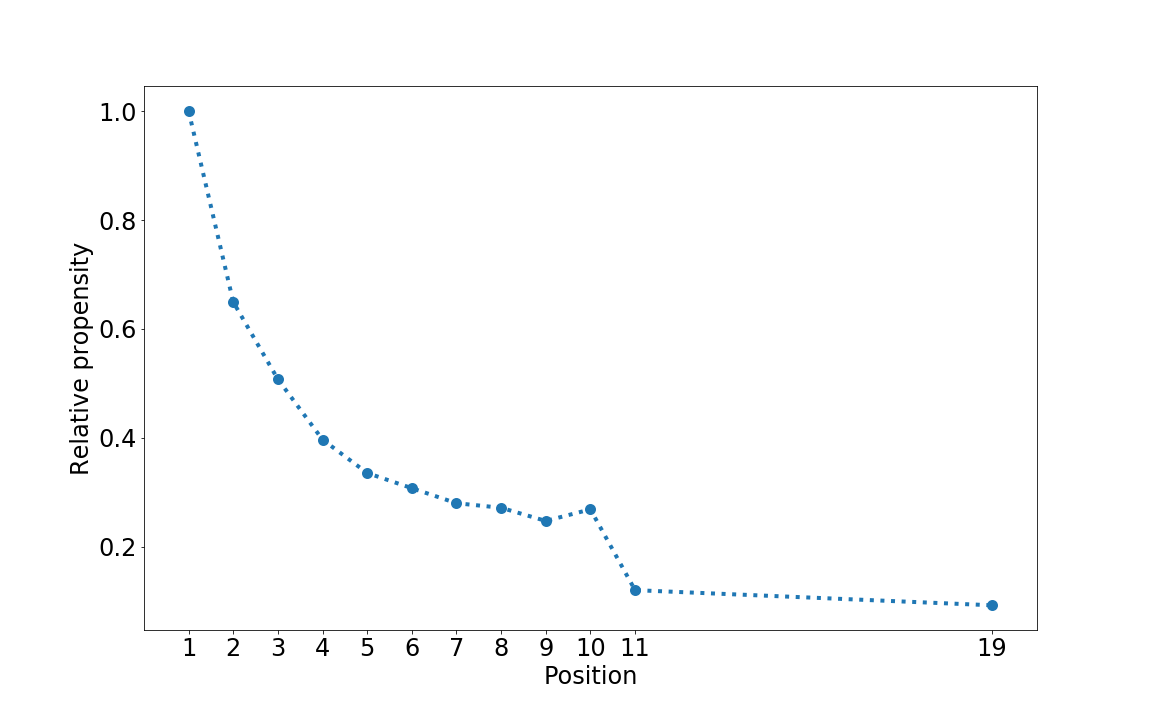}
  \caption{Position bias estimates from the RandPair algorithm.}
  \label{fig:position_bias_estimates}
\end{figure}

Results for our most common search result interface are shown in Figure \ref{fig:position_bias_estimates}. We see sharply decaying relative propensities in the first few positions of the search results. This bias tapers off as we progress down the search results. We observe a small bias in favour of position 10 over position 9, fitting with the higher contact rates we observe at the bottom of that page and at the bottom of subsequent pages. We see a small drop between positions 11 and 19, suggesting a very flat position bias curve between those positions.

\section{Successful applications}

Thumbtack ranks using predictive models based on observed historical relationships, in the class of machine learning models where algorithms use data to determine how values for each feature should impact a probability estimate. We call a comprehensive ranking algorithm, typically with machine learning models at its core, a \textit{ranker}.

Ultimately we evaluate rankers through visitor-randomized or market-randomized experiments, where we test new rankers ("challengers") against the existing ranker (the "baseline" or "champion"). While offline metrics for ranking are dependent on adjustments (or non-adjustments) for ranking bias, metrics from randomized experiments are a better test of real world performance. We use metrics of marketplace success, such as the rate of successful matches between customers and professionals. We have many experiments, and typically allocate enough sample to power them to support two-sided t-tests with a minimum detectable effect of a 1\% improvement at significance of 0.05 and power of 80\%. While occasionally we test individual features in isolation, more commonly we incorporate multiple changes into a single new ranker.

Following our randomization program our next two successful challengers both made substantial use of our calculated position bias factors, with several unrelated unsuccessful challengers in between. These rankers are discussed below.

\subsection{Historical rate features}

We can create a ranking of professionals because those professionals differ on qualities intrinsic to each professional, dynamic in the marketplace, and relative to the search.

A particular class of dynamic features are those based on historical aggregate customer behavior with respect to each professional. We calculate historical contact rates for each professional. Adopting notation from \citet{Chapelle09}, let there be $n$ searches in which professional $u$ appeared, with $C$ as a vector of binary contact outcomes. Then the historical contact rate for professional $u$ is $\alpha_u$, as calculated in Equation \ref{eq:raw_coec}.

\begin{equation}
  \alpha_u = \frac{\sum_{i=1}^{n} C_i}{n}
  \label{eq:raw_coec}
\end{equation}

Historical rates are powerful because they embed causal information that may be hard to fully enumerate and quantify through other features. Customers may care about average ratings, number of reviews, enthusiasm within review text, business names, photos, key terms in the business profile, and anything else they can observe on Thumbtack. Thumbtack may know much of this information, but at any point we will have less than perfect identification of all possible features. Historical contact rates could reflect information that Thumbtack has yet to identify and quantify. Our ranking models use historical rates as input features, alongside many other features.

The existence of the ranking problem implies that ranking bias exists, and witnessing that bias has motivated a substantial literature. Otherwise we would be unable to influence customer outcomes through ranking algorithms. Historical rates depend on customer behavior and are thus directly affected by ranking bias. Consider two professionals regularly shown in the same searches, one consistently in the first position and the other in the 15th position. The first professional will almost certainly receive more contacts than the latter even if they both have similar quality.

If historical rates as in Equation \ref{eq:raw_coec} were significant positive factors in our models then the rankings could be self-perpetuating, with the top ranked professionals cementing their position and the lowest ranked professionals facing a severe uphill battle to ever displace higher professionals. This self-perpetuating ranking results in the popularity bias problem, with a very active literature such as recent publications by \citet{Wei21, Zhang21}, and \citet{Zhu21}. The use of historical rate features compounds the importance of estimating ranking bias.

The clicks over expected clicks (COEC) model adjusts each professional’s historical metric relative to the average at their position \citep{Chapelle09, Ling17}. Each interaction can be weighed by the inverse of the propensity of contact at that position. For example, if the top position is contacted 10\% of the time and the 15th position 1\% of the time, then we could give 10 times the credit to a contact at position 15 as a contact at position 1. Two professionals can be ordered not by their absolute rate of contacts per search, but their rate of contacts per search relative to their position. A professional who is above average for their position will have a higher score than a professional who is below average at their position, regardless of what those positions are. Equation \ref{eq:coec} shows the updated calculation, with $\theta$ representing the vector of position weights and $k$ the vector of positions in which the professional is shown.

\begin{equation}
  \alpha_u = \frac{\sum_{i=1}^{n} C_i}{\sum_{i=1}^{n} \theta_{k_i}}
  \label{eq:coec}
\end{equation}

The issue with this approach is that if we use non-randomized data to determine our $\theta$ position weights then the formula is not an accurate adjustment for position bias. Professionals in the highest positions are contacted more often than professionals in lower positions not only due to the causal impact of rankings on searcher behavior, but also because we intentionally try to rank the best matches in the highest positions. If we try to measure position bias by observed interaction rates we conflate ranking bias with match quality and will overestimate the position bias. If rankings were entirely random then contact rates by position would be an unbiased estimate of position bias, but our core goal is to improve rankings and keep them quite different from random.

For some time we used this type of adjusted historical rates as features in our models. The position weights were derived from our earliest ranking data, when our ranking algorithm was in its infancy and less correlated with match quality than later iterations. We knew the feature contained a quality component, likely overstating position bias. This could create a small countervailing reversion to the mean effect, where low-ranked professionals were over-credited for their interactions and high ranked professionals faced an unfair challenge in maintaining their position, since our ranking models had a monotonically positive relationship between these COEC features and predicted positive outcomes. 

Whether biased COEC features are a problem depends on the impact of the bias on our ranking models. If we wanted to predict contact as a function of current position, with position being constant, then raw historical rates could be strong features. However, in a context where we optimize the ranking itself, bias that is not predictive of position-invariant match quality may limit the usefulness of these features in models. 

As an outcome of our randomization program we created new versions of our historical rate features, using the position bias factors shown in Figure \ref{fig:position_bias_estimates}. In subsequent models we found the new features to be stronger, as measured by feature importance metrics, and they largely displaced the older versions of rate features. We conducted a visitor-randomized experiment on live traffic comparing our baseline against a new ranking model which included these updated COEC features. The baseline used the older (biased) version of rate features. We included several other new features in the new model, unfortunately entangling the effects of multiple changes, but the other new features were among our most marginal by feature importance metrics while the COEC features are among our strongest. The ranker became our new baseline after substantial and statistically significant improvements, as shown in Table \ref{table:improvedratefeatures}. In many subsequent experiments the updated COEC features continue to be among our strongest features and continue to have more predictive power than their predecessors.

\begin{table}
  \centering
  \caption{Improvement of the unbiased COEC model over the baseline in a 50/50 visitor-randomized A/B test}
  \begin{tabular}{ccl}
    \toprule
    Metric&Performance&p-value\\
    \midrule
    Rate of visitors with >= 1 contact & +1.6\% & 0.01\\
    Rate of visitors with >= 1 match & +1.7\% & 0.01\\
  \bottomrule
\end{tabular}
\label{table:improvedratefeatures}
\end{table}

\subsection{Ranking models under the PBM}

After identifying position bias factors and transitioning to models using new historical rate features, we built a new ranker using Inverse Propensity Scoring (IPS). Applying notation from \citet{Vardasbi20}, Equation \ref{eq:ips} describes how the estimated loss ($\tilde{\Delta}$) of a ranker ($f$) is the sum across a corpus of $n$ queries of all contacts on each document ($C_i(d)$) multiplied by a (typically) position-based weighting ($\lambda$) for that document divided by a position bias factor ($\theta_k$). The $\lambda$ function could be the position of the document shown (creating the Average Relevance Position (ARP), as used in \citet{Joachims17}) or position could have a logarithmic penalty using Discounted Cumulative Gain (DCG), among other options.

\begin{equation}
  \tilde{\Delta}_{IPS}(f) = \frac{1}{N}\sum_{i=1}^{n}\sum_{(d,k) \in y_i} \frac{C_i(d)}{\theta_k} \cdot \lambda(d|q_i, f)
  \label{eq:ips}
\end{equation}

In the example case of using DCG as the $\lambda$ loss metric, the only modification to traditional DCG is the division by position bias weights ($\theta_k$). IPS rewards models that would rank the contacted professionals at higher positions, weighing observations more when they have less favourable position bias.

IPS is an ideal loss functions under the assumptions of the PBM \citep{Joachims17}. The separability assumption is key, and in the presence of trust bias that varies by position the IPS loss function will not be optimal \citep{Agarwal19b, Vardasbi20}. We chose to build an IPS model even with the risk of trust bias or other model misspecification, because of the simplicity of model estimation, the literature of successful IPS uses, and strong existing implementations. 

We found that our IPS iteration improved our marketplace metrics in a visitor-randomized experiment by a meaningful amount over our baseline at the time, which was the model with unbiased historical features discussed earlier. Table \ref{table:ipsmodel} shows the results. The IPS model became our new baseline, and we continued to use IPS in subsequent iterations.

\begin{table}
  \centering
  \caption{Improvement of the IPS model over the unbiased COEC model in a 50/50 visitor-randomized A/B test}
  \begin{tabular}{ccl}
    \toprule
    Metric&Performance&p-value\\
    \midrule
    Rate of visitors with >= 1 contact & +0.7\% & 0.08\\
    Rate of visitors with >= 1 match & +0.9\% & 0.05\\
  \bottomrule
\end{tabular}
\label{table:ipsmodel}
\end{table}

Along with the loss function change we incorporated several new features, since we continually develop new features for subsequent model experiments. Unfortunately this permits multiple hypotheses of improved model performance. Improvements could be due to IPS, could be due to new features, or could be caused by model drift \citep{Gama14}. While we tested isolating changes in offline evaluation that indicated improvements separately from IPS and from new features, we believe that experimenting on users is a more complete test of improvement. Given finite volume of visitors to test new rankers on, and that earlier application of improvements leads to more cumulative value than delayed improvements, we often use models that incorporate multiple changes.

Further observational evidence supporting the IPS approach is that it has survived challenges since, including tests of a listwise model with LambdaMART and a trust bias model using the implementation from \citet{Vardasbi20}.

\subsection{Internal and external tooling}

With position bias estimates available we are better able to answer hypothetical questions about how alternative rankers would affect our customer metrics. These scenarios helped us make decisions on product changes. The parsimony of randomization helped to explain the results, making it easy to build trust in our applications.

In a pilot program we provided professionals with a tool to understand how their ranking and contacts could be affected as they change some of the parameters of their business. This tool used position bias estimates to forecast how ranking changes would likely affect their volume of contacts.

Neither of these applications were the original motivation for the randomization program. We expect that having a better understanding of ranking bias and customer behavior will continue to help us through future new ideas.

\section{Conclusion}

This paper summarizes the choice of ranking bias model, measurement of position bias, and applications at a large consumer marketplace. After due consideration of several behavior models, we chose to apply a position-based propensity model. Applying a RandPair swap intervention program, we discuss sampling and calculation considerations. We describe how position bias can influence historical rate features in machine learning models, improving model performance even before adopting inverse propensity scoring as a loss function for model optimization. Our position bias estimates directly led to improved ranking and better scenario tooling.

\section*{Acknowledgements}
We thank Thumbtack colleagues for creating a vibrant intellectual environment where this work could have a large positive impact on our marketplace, and all members of the Marketplace Matching team for working on this and other initiatives. In particular we thank Mark Andrew Yao for sharing \citet{Wang18}, Renal Khabibulin for sharing \citet{Li20}, and Amber Wang for proposing and eventually creating the IPS ranker.

\bibliographystyle{plainnat}
\bibliography{references}

\end{document}